\documentclass[%
 reprint,
 longbibliography,
 amsmath,amssymb,
 aps,
]{revtex4-1}

\usepackage{graphicx}
\usepackage{color}
\usepackage{placeins}
\usepackage{float}
\usepackage{hyperref}
\usepackage{tabularx,colortbl}
\usepackage{amsmath}
\usepackage{bm}
\usepackage{caption}
\usepackage{subcaption}
\usepackage{dcolumn}% Align table columns on decimal point
\usepackage{color}
\usepackage{multirow}
\usepackage[normalem]{ulem}
\usepackage[mathlines]{lineno}% Enable numbering of text and display math
\begin{document}

\preprint{APS/123-QED}

\title{Design and experimental demonstration of impedance-matched circular polarization selective surfaces with spin-selective phase modulations}

\author{Minseok Kim and George V. Eleftheriades}
\email{gelefth@ece.utoronto.ca}
\affiliation{%
The Edward S. Rogers Department of Electrical and Computer Engineering, University of Toronto, Toronto, Canada
}%

\date{\today}

\begin{abstract}
This paper presents the design and experimental demonstration of an impedance-matched circular polarization selective surface which also offers spin-selective phase modulations at microwave frequencies. We achieve this by leveraging the theory of Pancharatnam-Berry phase and cascading four tensor impedance layers, each comprising an array of crossed meander lines. These meander lines are precisely tuned and rotated to implement particular tensor surface impedance values to satisfy the impedance-matching condition for the transmitted right-handed circularly-polarized field while inducing Pancharatnam-Berry phase shift for the reflected left-handed circularly-polarized field. We present a detailed numerical synthesis technique to obtain the required impedance values for satisfying the impedance matching condition, and demonstrate spin-selective phase modulations based on Pancharatnam-Berry phase shifts. To verify the proposed idea, we experimentally demonstrate nearly-reflectionless transmission of right-handed circular polarization at broadside and reflection of left-handed circular polarization at 30$^\circ$ off broadside \textcolor{black}{at 12 GHz}. \textcolor{black}{For this purpose}, a free-space quasi-optical set up and a near-field measurement system are respectively employed for measuring the transmitted and reflected circularly-polarized fields.
%\begin{description}
%\item[PACS numbers]
%\centering
%41.20.Jb, 73.20.Mf, 78.67.Pt, 71.45.Gm
%\end{description}
\end{abstract}

\maketitle

\section{\label{sec:Intro}Introduction}

The idea of full control on the reflective and transmissive properties of electromagnetic (EM) waves has triggered a huge research interest from both \textcolor{black}{the} physics and engineering communities due to its numerous potential applications. In this regard, recent years have witnessed a rapid development in the field of metasurfaces which are artificial EM surfaces comprising arrays of subwavelength-sized unit cells. By engineering the local interaction properties between the unit cells and an incident EM field, metasurfaces for various functionalities have been proposed to date~\cite{Yu2011,Selvanayagam2013OptEx,Yu2013,Monticone2013,Kim2014PRX,Lin2014,SemchenkoEJAP2010,Wong2016IEEE}. For example, phase-gradient metasurfaces have been demonstrated for functionalities such as light bending~\cite{Yu2011,Yu2013,Ni2012,Monticone2013,Selvanayagam2013OptEx}, holograms~\cite{Ni2013,Huang2013,Kim2016JOSAB}, and orbital angular momentum generation~\cite{Chen2015NanoLett,Bi2018OptExp,Ren2019NatComm}. In these phase-gradient metasurfaces, unit cells are tailored to provide certain reflection/transmission phase profiles along the surfaces such that they locally match the tangential wave-vectors for anomalously reflected/refracted fields. As such, much of the effort has been \textcolor{black}{placed on} the demonstrations of unit cells that provide a full range of 360$^\circ$ of transmission/reflection phase shifts. In parallel to these works, those that arbitrarily control the polarization state of the scattered waves have also been extensively studied. For example, numerous metasurfaces that leverage birefringence have been reported for functionalities such as quarter-wave and half-wave plates~\cite{Yu2012NanoLett,Pors2013OptExA,Lin2014,Wu2016APL}. These birefringent metasurfaces, however, have limited polarization control of EM waves in the sense that they fail to control the flow (e.g., phase velocity) of different circular polarizations (CPs). To also control the flow of both left-handed circular polarization (LHCP) and right-handed circular polarization (RHCP), chiral metasurfaces have been demonstrated which involve the utilization of uniform bianisotropic unit cells. Amongst various chiral polarization transformations, CP selectivity has gained particular interest owing to its promising applications in areas such as satellite communications~\cite{Sanz2015IEEE}. In particular, Zhao \textit{et al.} have captured much attention with their proposal on a `twisted-metamaterials' for realizing a circular polarization selective surface (CPSS) which transmits one handedness of a CP field while reflecting the opposite handedness~\cite{Zhao2012NatCommun}. The proposed multi-layer scheme consists of cascaded layers of identical unit cells that are progressively rotated along the propagation direction of an incident field. In this respect, their operation is akin to mimicking a helical structure which is a known geometry for chiral molecules found in nature~\cite{Zhao2012NatCommun,Ericsson2017IEEE,Wang2016IOP}. As such, they are not necessarily optimal in the sense that they are not guaranteed to be impedance matched, which can lead to degraded efficiency from undesired reflections, because they merely rely on a rotated lattice effect to naturally realize chirality. On the other hand, subsequent works by Selvanayagam \textit{et al.} have demonstrated an impedance-matched CPSS by cascading two surfaces with both electric and magnetic polarizations~\cite{Selvanayagam2014IEEE}, and cascading three surfaces with only electric polarizations~\cite{Selvanayagam2016IEEE}. Following these works, Kim \textit{et al.} have also demonstrated a similar impedance-matched CPSS that operates at two user-defined frequency bands by cascading dual-resonance tensor impedance surfaces~\cite{Kim2018SciRep}.

Although these previous CPSSs have been successfully demonstrated, they commonly utilize uniform unit-cell arrays. As a result, other than a polarization transformation, they cannot perform arbitrary beam-shaping for the reflected/transmitted CP fields. However, a CPSS which also combines the functionalities of phase-gradient metasurfaces is of great interest in many applications (e.g., multiple beam formation in satellite communications). In this regard, researchers have recently started to report a few chiral metasurfaces that can also re-shape the reflected/transmitted CP fields. For example, at the cost of increased complexity in fabrication, a curved CPSS has been demonstrated in \cite{Cappellin2016IEEE} where the CPSS is physically shaped to obtain the necessary reflection phase shifts from the wave propagation. On the other hand, planar phase-gradient chiral metasurfaces have also been demonstrated that take advantage of the geometrical phase shift~\cite{Zhou2016EuCap,Jing2017LPR,Chen2018LSA,Ma2018OptExp}. In particular, these phase-gradient chiral metasurfaces realize spin-selective phase modulations by rotating individual chiral unit cells to acquire the Pancharatnam-Berry phase shifts along their surfaces~\cite{Phelan1976MicrowaveJournal}. Based on this principle, \cite{Jing2017LPR} has demonstrated absorption for one handedness of a CP field and arbitrary reflection for the other handedness. Another phase-gradient CPSS has been demonstrated in \cite{Ma2018OptExp} which normally reflects an incident LHCP field while applying arbitrary transmission phase shift for an incident RHCP field. Despite their successful demonstrations, however, none of these previous works has shown a systematic way to also satisfy the impedance-matching condition for the transmitted CP fields, while maximizing the reflection for the opposite handedness. In this sense, they do not possess optimal efficiency as they do not offer a \textcolor{black}{means} to suppress any unwanted reflections or transmissions for their surfaces.

In departure from the aforementioned previous works, we hereby propose and experimentally demonstrate a CPSS that is precisely designed to (a) satisfy the impedance-matching condition for the transmitted RHCP field and (b) induce spin-selective phase modulations for the reflected LHCP field. We achieve this by leveraging the theory of Pancharatnam-Berry phase shift and cascading four tensor impedance layers in which each layer precisely implements particular impedance values. Whereas the preliminary numerical results have been reported in \cite{Kim2019IEEEAPS}, this study shows the detailed numerical synthesis method for obtaining the required impedance values by modeling the four-layered system in a multi-conductor transmission line. The computed surface impedances are then physically encoded by utilizing rotated crossed meander lines as our unit cells at microwave frequencies (12 GHz). As it shall be shown, the proposed unit cell can implement a large range of either capacitance or inductance for two orthogonal linearly-polarized (LP) waves. Moreover, the unit cell does not rely on any resonances for accessing these capacitances or inductances thereby minimizing Ohmic losses. Based on the proposed unit cell, we numerically and experimentally demonstrate a case where an incident LHCP field is fully reflected at a certain prescribed angle while transmitting an incident RHCP field with minimum reflection. In particular, we use a free-space quasi-optical system for measuring the transmission of an RHCP field and the near-field measurement system for verifying anomalous reflection of an LHCP field. \textcolor{black}{The demonstrated example is of particular interest in many applications such as CP detectors, polarizers, spin-selective orbital angular momentum generation, and multiple beam generation for satellite communications.}

\section{Design of an impedance-matched CPSS with spin-selective phase modulations}

\begin{figure*}[!ht]
\centering
  \captionsetup[subfigure]{justification=centering}
  \includegraphics[width=\textwidth]{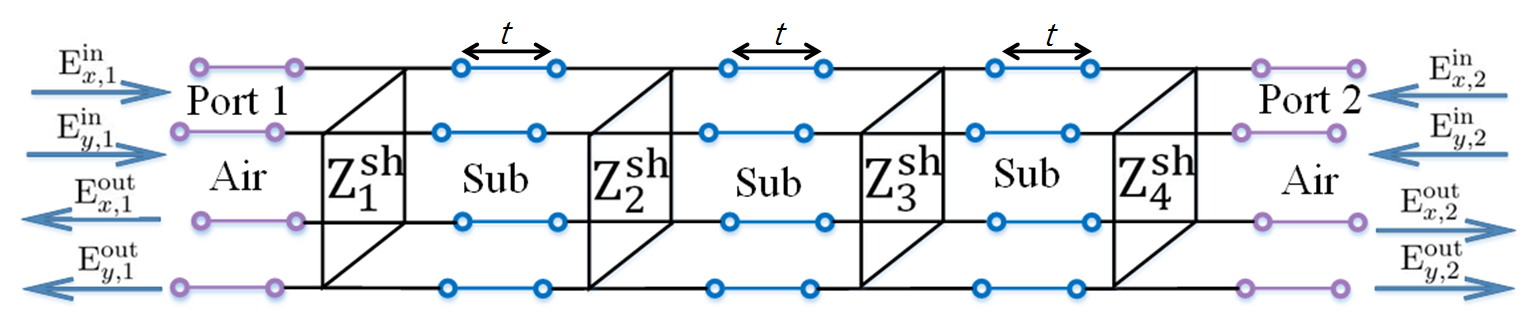}
  \caption{The schematic of the impedance-matched CPSS in a multi-conductor transmission line system. The system effectively consists of four ports: two for the physical ports and the other two for the orthogonal polarization states. The superscripts ``in'' and ``out'' respectively denote whether a field is an input or an output. The subscripts $x$ and $y$ represent the polarization states of the fields. Specifically, $x$ and $y$ respectively represent LP fields that are polarized along the $x$- and $y$-directions. The numbers, 1 and 2, in the subscript denote the physical port numbers.}
  \label{fig:SuperCellOverview}
\end{figure*}

The objective of this paper is to realize an ideal CPSS that is (a) impedance-matched and (b) offers spin-selective phase modulations. To this end, we consider cascading four impedance layers similar to our previous works in Refs.~\cite{Selvanayagam2016IEEE,Kim2016OL,Kim2018SciRep} which have demonstrated uniform chiral metasurfaces (i.e., a constant magnitude and phase for the reflected/transmitted CP fields). While different number of layers can be considered (minimum of three layers as explained in \cite{Selvanayagam2016IEEE}), we hereby utilize an additional impedance layer to add a degree of freedom. The schematic of the proposed cascaded impedance layers is shown in Fig.~\ref{fig:SuperCellOverview} where each layer is represented by an ideal tensor impedance surface in \textcolor{black}{a} multi-conductor transmission line system.
In what follows, \textcolor{black}{Subsection \ref{SubsectionIIA}} discusses the numerical and physical designs of an impedance-matched CPSS based on the proposed multi-layered scheme. The devised impedance-matched CPSS is then later utilized as a building block for demonstrating spin-selective phase modulations in the following subsection (\textcolor{black}{Section \ref{SubsectionIIB}}).

\subsection{Synthesis of an impedance-matched CPSS}\label{SubsectionIIA}
To realize an impedance-matched CPSS based on the proposed four cascaded tensor impedance layers shown in Fig.~\ref{fig:SuperCellOverview}, we first envision its net scattering matrix, ${\mathbf{S}_{\text{UC}}}$, as
\begin{equation}
\begin{aligned}
{\mathbf{S}_{\text{UC}}} &=
\begin{bmatrix}
\text{S}_{11}^{xx} & \text{S}_{11}^{xy} & \text{S}_{12}^{xx} & \text{S}_{12}^{xy}\\ 
\text{S}_{11}^{yx} & \text{S}_{11}^{yy} & \text{S}_{12}^{yx} & \text{S}_{12}^{yy}\\ 
\text{S}_{21}^{xx} & \text{S}_{21}^{xy} & \text{S}_{22}^{xx} & \text{S}_{22}^{xy}\\ 
\text{S}_{21}^{yx} & \text{S}_{21}^{yy} & \text{S}_{22}^{yx} & \text{S}_{22}^{yy}
\end{bmatrix} = \frac{1}{2}\begin{bmatrix} -1 & j & 1 & -j \\ j & 1 & j & 1 \\ 1 & j & -1 & -j \\ -j & 1 & -j & 1 \end{bmatrix}e^{j\xi},
\end{aligned}
\label{eq:SCPSSDef}
\end{equation}
\noindent where $\xi$ represents an arbitrary phase constant. The scattering matrix in \eqref{eq:SCPSSDef} is in an LP basis and it is defined as (referring to Fig.~\ref{fig:SuperCellOverview}),
\begin{equation}
\begin{bmatrix}
\text{E}_{x,1}^{\text{out}} \\ 
\text{E}_{y,1}^{\text{out}} \\ 
\text{E}_{x,2}^{\text{out}} \\ 
\text{E}_{y,2}^{\text{out}}
\end{bmatrix}
=
\begin{bmatrix}
\text{S}_{11}^{xx} & \text{S}_{11}^{xy} & \text{S}_{12}^{xx} & \text{S}_{12}^{xy}\\ 
\text{S}_{11}^{yx} & \text{S}_{11}^{yy} & \text{S}_{12}^{yx} & \text{S}_{12}^{yy}\\ 
\text{S}_{21}^{xx} & \text{S}_{21}^{xy} & \text{S}_{22}^{xx} & \text{S}_{22}^{xy}\\ 
\text{S}_{21}^{yx} & \text{S}_{21}^{yy} & \text{S}_{22}^{yx} & \text{S}_{22}^{yy}
\end{bmatrix}
\begin{bmatrix}
\text{E}_{x,1}^{\text{in}} \\ 
\text{E}_{y,1}^{\text{in}} \\ 
\text{E}_{x,2}^{\text{in}} \\ 
\text{E}_{y,2}^{\text{in}}
\end{bmatrix},
\label{eq:SMatLPGen}
\end{equation}
\noindent where the scattering parameters in an LP basis are defined as,
\begin{equation}
\text{S}_{ij}^{\alpha \beta} = \left. \frac{\text{E}_{\alpha,i}^{\text{out}}}{\text{E}_{\beta,j}^{\text{in}}} \right]_{\text{E}_{\kappa\neq\beta,l}^{\text{in}} = 0}.
\label{eq:SparamLP}
\end{equation}
\noindent In \eqref{eq:SparamLP}, $\alpha$ and $\beta$ represent the direction of linear polarizations (i.e., either $x$ or $y$), or equivalently different propagating modes at the output port (port $i$) and the input port (port $j$), respectively. The condition, $\text{E}_{\kappa\neq\beta,l}^{\text{in}} = 0$, ensures the only input to be $\text{E}_{\beta,j}^{\text{in}}$.

By multiplying \eqref{eq:SCPSSDef} with the Jones matrices for an incident LHCP field and RHCP field from port 1 or 2, it can be readily shown that the transmission magnitude of an RHCP field is identically 1 (i.e., impedance-matched and lossless), while that of an LHCP field is identically 0 (i.e., perfect reflection). Furthermore, the axial ratio for the transmitted RHCP field and the reflected LHCP field ideally remain unity, which implies that an RHCP field is transmitted into a pure RHCP field, while an LHCP field is reflected as a pure LHCP field. Therefore, ${\mathbf{S}_{\text{UC}}}$ in \eqref{eq:SCPSSDef} is our desired net scattering matrix for the cascaded system and the goal here is to obtain the required surface impedance values in each layer such that they cascade to match to ${\mathbf{S}_{\text{UC}}}$.

To obtain the required surface impedance values, we first modularize each component in the proposed four-layered system shown in Fig.~\ref{fig:SuperCellOverview}. In particular, we modularize the system into two main parts: (1) the transmission-line module and (2) tensor impedance layer module. The first module describes the wave propagation between the tensor impedance layers, and it can be mathematically represented by its scattering matrix, $\mathbf{S}_{\text{TL}}$, which is given as,
\begin{equation}
\label{eq:STLDef}
{\mathbf{S}_{\text{TL}}} = \begin{bmatrix} 0 & 0 & e^{-jk_\text{o}tn_{\text{sub}}} & 0 \\ 0 & 0 & 0 & e^{-jk_\text{o}tn_{\text{sub}}} \\ e^{-jk_\text{o}tn_{\text{sub}}} & 0 & 0 & 0 \\ 0 & e^{-jk_\text{o}tn_{\text{sub}}} & 0 & 0 \end{bmatrix},
\end{equation}
\noindent where $k_\text{o}$, $t$, and $n_{\text{sub}}$ respectively represent the wave-vector, separation length, and refractive index of the material between the impedance layers. For the remainder of this work, we assume that $t$ is fixed to 3.175mm and the material between the layers is \textcolor{black}{a} Rogers 5880 substrate (dielectric permittivity of 2.2). On the other hand, the second module that describes the $n^{\text{th}}$ tensor impedance layer can also be represented by its scattering matrix, $\mathbf{S}_{\text{n}}$, which is given as,
\begin{equation}
\mathbf{S}_{n} = \mathbf{G}_{\text{ref}} \cdot \left ( \mathbf{Z}_{n}^{\text{sh}} - \mathbf{Z}_{\text{ref}} \right ) \cdot \left ( \mathbf{Z}_{n}^{\text{sh}} + \mathbf{Z}_{\text{ref}} \right )^{-1} \cdot \mathbf{G}_{\text{ref}}^{-1},
\label{eq:ZtoS}
\end{equation}
\noindent where $\mathbf{Z}_{\text{ref}}$ and $\mathbf{G}_{\text{ref}}$ are diagonal matrices whose non-zero components are the reference port impedances and normalized port admittances, respectively. $\mathbf{Z}_{n}^{\text{sh}}$ represents the impedance matrix for the $n^{\text{th}}$ tensor impedance layer and it is defined as,
\begin{equation}
\mathbf{Z}_{n}^{\text{sh}} =  \begin{bmatrix}
\text{Z}_{n}^{xx} & \text{Z}_{n}^{xy} & \text{Z}_{n}^{xx} & \text{Z}_{n}^{xy} \\ \text{Z}_{n}^{yx} & \text{Z}_{n}^{yy} & \text{Z}_{n}^{yx} & \text{Z}_{n}^{yy} \\ \text{Z}_{n}^{xx} & \text{Z}_{n}^{xy} & \text{Z}_{n}^{xx} & \text{Z}_{n}^{xy} \\ \text{Z}_{n}^{yx} & \text{Z}_{n}^{yy} & \text{Z}_{n}^{yx} & \text{Z}_{n}^{yy}
\end{bmatrix}.
\label{eq:ZMTX}
\end{equation}
\noindent \textcolor{black}{Here, the impedance parameters in \eqref{eq:ZMTX} relate the difference in the magnetic fields at the boundary formed by the $n^{\text{th}}$ tensor impedance layer to the continuous tangential electric field at the boundary. In other words, the impedance matrix given in \eqref{eq:ZMTX} describes the ratio between the induced surface currents on the $n^{\text{th}}$ impedance layer and the tangential electric components of the incident and scattered fields on the $n^{\text{th}}$ layer. Hence, it is given by~\cite{Selvanayagam2016IEEE}}
\begin{equation}
 \begin{bmatrix} Ex \\ Ey \end{bmatrix} =
 \begin{bmatrix} \text{Z}_{n}^{xx} & \text{Z}_{n}^{xy} \\ \text{Z}_{n}^{yx} & \text{Z}_{n}^{yy} \end{bmatrix}
 \begin{bmatrix} -(H_y^+-H_y^-) \\ H_x^+-H_x^- \end{bmatrix}.
\label{eq:ZParamRelation}
\end{equation}
\noindent These impedance parameters can be directly extracted from ANSYS HFSS simulation\textcolor{black}{s} by simulating the $n^{\text{th}}$ layer in a periodic boundary condition with Floquet port excitation. Specifically, the relation is given as,
\begin{equation}
\text{Z}_{n}^{\alpha\beta} = \text{Z}_{\text{HFSS}}[2,\alpha,1,\beta],
\label{eq:ZHFSStoZserZsh}
\end{equation}
\noindent where $\text{Z}_{\text{HFSS}}$ denotes the impedance parameter which the HFSS simulation computes and it is in the form of $\text{Z}_{\text{HFSS}}$[output port $i$, output mode $\alpha$, input port $j$, input mode $\beta$]. The 2$\times$2 impedance matrix in \eqref{eq:ZParamRelation} can also be diagonalized as,
\begin{equation}
 \begin{bmatrix}
 \text{Z}_{n}^{xx} & \text{Z}_{n}^{xy} \\ \text{Z}_{n}^{yx} & \text{Z}_{n}^{yy}
 \end{bmatrix} =
 \mathbf{R}(\theta)
 \begin{bmatrix}
 \text{Z}_{n}^{X} & 0 \\ 0 & \text{Z}_{n}^{Y} \end{bmatrix}
 \mathbf{R}^{-1}(\theta),
\label{eq:ZParamRelationDiag}
\end{equation}
\noindent where $\text{Z}_{n}^{X}$ and $\text{Z}_{n}^{Y}$ are the eigenvalues of the 2$\times$2 impedance matrix in \eqref{eq:ZParamRelation} and $\mathbf{R}$ is a square matrix whose columns are the linearly independent eigenvectors of the 2$\times$2 impedance matrix. Physically speaking, $\mathbf{R}$ is a 2$\times$2 rotational matrix and $\theta$ is the angle for which the whole layer is rotated by. Therefore, it is of the form given by
\begin{equation}
\mathbf{R}(\theta) = \begin{bmatrix}
\text{cos}(\theta) & -\text{sin}(\theta)\\ 
\text{sin}(\theta) & \text{cos}(\theta)
\end{bmatrix}.
\label{eq:R2by2}
\end{equation}

Provided that the impedance parameters in each layer are known, the scattering matrices that describe each module can be constructed based on \eqref{eq:STLDef} and \eqref{eq:ZtoS}. With all these scattering matrices, the net scattering matrix of the overall system can then be obtained by cascading them via the generalized scattering matrix (GSM) method. For example, two modules which are represented as $\mathbf{S}^{\text{E}}$ and $\mathbf{S}^{\text{F}}$ can be cascaded as,
\begin{subequations}
\begin{align}
\mathbf{S}_{11}^{\text{EF}} &= \mathbf{S}_{11}^{\text{E}} + \mathbf{S}_{12}^{\text{E}}\left(\mathbf{I}-\mathbf{S}_{11}^{\text{F}}\mathbf{S}_{22}^{\text{E}}\right)^{-1}\mathbf{S}_{11}^{\text{F}}\mathbf{S}_{21}^{\text{E}}\\
\mathbf{S}_{12}^{\text{EF}} &= \mathbf{S}_{12}^{\text{E}}\left(\mathbf{I}-\mathbf{S}_{11}^{\text{F}}\mathbf{S}_{22}^{\text{E}}\right)^{-1}\mathbf{S}_{12}^{\text{F}}\\
\mathbf{S}_{21}^{\text{EF}} &= \mathbf{S}_{21}^{\text{F}}\left(\mathbf{I}-\mathbf{S}_{22}^{\text{E}}\mathbf{S}_{11}^{\text{F}}\right)^{-1}\mathbf{S}_{21}^{\text{E}}\\
\mathbf{S}_{22}^{\text{EF}} &= \mathbf{S}_{22}^{\text{B}} + \mathbf{S}_{21}^{\text{F}}\left(\mathbf{I}-\mathbf{S}_{22}^{\text{E}}\mathbf{S}_{11}^{\text{F}}\right)^{-1}\mathbf{S}_{22}^{\text{E}}\mathbf{S}_{12}^{\text{F}}
\end{align}
\end{subequations}
\noindent which arise by taking proper account for all reflected and transmitted field/voltage vectors between the two modules~\cite{Bhattacharyya2006}. Here, $\mathbf{S}_{11}^{\text{EF}}$, $\mathbf{S}_{12}^{\text{EF}}$, $\mathbf{S}_{21}^{\text{EF}}$, and $\mathbf{S}_{22}^{\text{EF}}$ are 2$\times$2 sub-matrices that comprise the net scattering matrix, $\mathbf{S}^{\text{EF}}$, which is in the form given as,
\begin{equation}
\begin{aligned}
{\mathbf{S}^{\text{EF}}} &=
\begin{bmatrix}
\text{S}_{11}^{xx} & \text{S}_{11}^{xy} & \text{S}_{12}^{xx} & \text{S}_{12}^{xy}\\ 
\text{S}_{11}^{yx} & \text{S}_{11}^{yy} & \text{S}_{12}^{yx} & \text{S}_{12}^{yy}\\ 
\text{S}_{21}^{xx} & \text{S}_{21}^{xy} & \text{S}_{22}^{xx} & \text{S}_{22}^{xy}\\ 
\text{S}_{21}^{yx} & \text{S}_{21}^{yy} & \text{S}_{22}^{yx} & \text{S}_{22}^{yy}
\end{bmatrix} = \begin{bmatrix}
\mathbf{S}_{11}^{\text{EF}} & \mathbf{S}_{12}^{\text{EF}} \\ 
\mathbf{S}_{21}^{\text{EF}} & \mathbf{S}_{22}^{\text{EF}}
\end{bmatrix}.
\end{aligned}
\label{eq:GSM_Def}
\end{equation}

The discussion thus far has demonstrated how the net scattering matrix of the cascaded layers can be obtained by modularizing each component to its corresponding scattering matrix. However, we have not yet specified the required impedance values in each layer which would cascade to result to the desired net scattering matrix in \eqref{eq:SCPSSDef}. Although there is no closed form solution for such a problem, there are several semi-analytical methods to obtain these impedance values. For example, one can solve the algebraic-Ricatti equation as demonstrated in \cite{Selvanayagam2016IEEE} which is similar to an impedance-matching problem or utilize the nonlinear optimization method as outlined in \cite{Kim2016OL,Kim2018SciRep}. The two methods are equally valid and we employ the latter method to solve the problem. Specifically, we utilize MATLAB's built-in optimizer, \texttt{fmincon}, to find a minimum of the cost function which is defined as,
\begin{equation}
\text{Cost} = \text{Max}(|{\mathbf{S}_{\text{net}}} - {\mathbf{S}_{\text{UC}}}|),
\label{eq:CostFunc}
\end{equation}
\noindent where $\mathbf{S}_{\text{net}}$ represents the net scattering matrix of the cascaded layers which involves varying all possible values of the impedance parameters in each layer. In optimizing $\mathbf{S}_{\text{net}}$, there are total 12 variables (2 eigenvalues and 1 rotation angle per each tensor impedance layer; see \eqref{eq:ZParamRelationDiag}). Nonetheless, we can dramatically reduce the number of variables by invoking the C$_4$ symmetry condition. Specifically, we set the eigenvalues of the first and last layers to be the same and their rotation angles to $+\theta_a$ and $-\theta_a$, respectively. Similarly, we force the eigenvalues of the second and third layers to be the same and their rotation angles to $+\theta_b$ and $-\theta_b$, respectively. Based on this scheme, the rotation angles and eigenvalues of the first and second layers are optimized at the operating frequency of 12 GHz such that the cost function is minimized to 0.0237. These optimized values are summarized in Table.~\ref{Tab:OptValues}
\begin{table}[]
\begin{tabular}{c|c|c|c|c|}
\cline{2-5}
 & Layer \#1 & Layer \#2 & Layer \#3 & Layer \#4 \\ \hline
\multicolumn{1}{|c|}{$\text{Z}^{X}$} & $j400\Omega$ & $-j256\Omega$ & $-j256\Omega$ & $j400\Omega$ \\ \hline
\multicolumn{1}{|c|}{$\text{Z}^{Y}$} & $-j240\Omega$ & $j40\Omega$ & $j40\Omega$ & $-j240\Omega$ \\ \hline
\multicolumn{1}{|c|}{$\theta$} & 64.4$^\circ$ & 18.5$^\circ$ & -18.5$^\circ$ & -64.4$^\circ$ \\ \hline
\end{tabular}
\captionsetup{justification=centering}
\caption{Optimized tensor impedance values}
\label{Tab:OptValues}
\end{table}

\begin{figure}[]
\centering
  \captionsetup[subfigure]{justification=centering}
  \begin{subfigure}{0.21\textwidth}
    \centering
    \includegraphics[width=\textwidth]{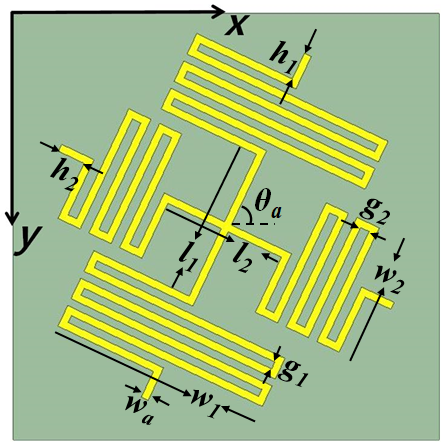}
    \caption{}
    \label{fig:LayerOne}
  \end{subfigure}
  \qquad
  \begin{subfigure}{0.215\textwidth}
    \centering
    \includegraphics[width=\textwidth]{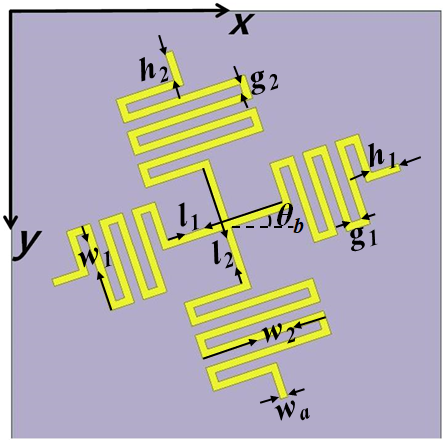}
    \caption{}
    \label{fig:LayerTwo}
  \end{subfigure}
  \begin{subfigure}{0.41\textwidth}
    \centering
    \includegraphics[width=\textwidth]{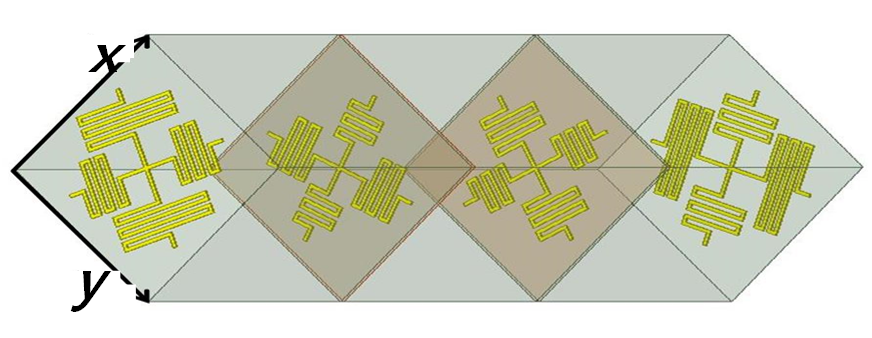}
    \caption{}
    \label{fig:CascadedStructure}
  \end{subfigure}
  \caption{The schematic of the proposed unit cells. (\subref{fig:LayerOne}) and (\subref{fig:LayerTwo}) The top views of the unit cells that implement the optimized tensor impedance values in the first and second layers and (\subref{fig:CascadedStructure}) the cascaded unit cells. We note that the first (or the second) and the last (or the third) layers are identical except that they are rotated in mirror symmetry}
  \label{fig:LayerOverview}
\end{figure}

\begin{figure}[]
\centering
\resizebox{0.5\textwidth}{!}{\includegraphics{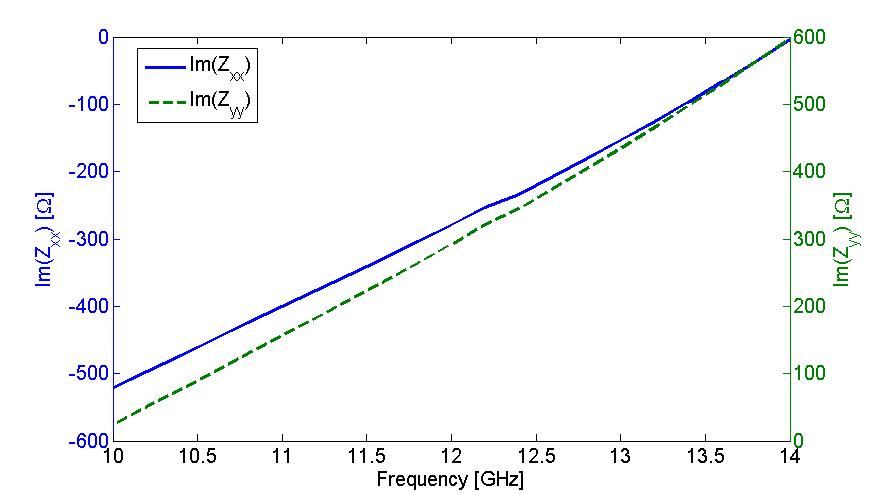}}
\caption{The frequency response of the proposed unit cell}
\label{fig:UCResponse}
\end{figure}

To verify and physically realize the optimized impedance values, we hereby propose crossed meander lines as our unit cells to form each layer. The top views of the proposed unit cells that comprise the first and second layers are respectively shown in Figs.~\ref{fig:LayerOne} and \ref{fig:LayerTwo}.
\noindent The unit-cell periodicity is 4.5 mm (${\sim}\lambda$/5.5) and they are printed on 3.175-mm-thick Rogers 5880 substrates. They are also bonded using 0.0508-mm-thick Rogers 2929 bondplies to form the proposed cascaded tensor impedance layers as shown in Fig.~\ref{fig:CascadedStructure} producing the overall thickness of ${\sim}$9.55 mm (${\sim}\lambda$/2.6). It should be noted that thinner substrates could have been used to realize more compact design; however, we have purposely employed thick substrates to obtain a sturdier structure. The rationale behind employing two orthogonal meander lines is to independently control the surface impedances for two orthogonal LP fields (i.e., $\text{Z}_{n}^{X}$ and $\text{Z}_{n}^{Y}$ in \eqref{eq:ZParamRelationDiag}). By changing the overall lengths of these meander lines and their number of turns, the unit cell can provide a wide range of capacitances and inductances for the two orthogonal fields. To demonstrate this point, Fig.~\ref{fig:UCResponse} shows the frequency response of the proposed unit cell near 12 GHz.
\noindent As seen, the unit-cell response for an $x$-polarized field (shown as a blue solid curve) is capacitive which ranges from $-j500$ $\Omega$ to $0$ $\Omega$ while that of an $y$-polarized field (shown as a green dotted curve) is inductive ranging from $0$ $\Omega$ to $j600$ $\Omega$. Because the unit cell offers an independent tuning of the surface impedances for the two orthogonal LP fields, with a wide range of capacitances and inductances, we can conveniently translate the required impedance values in Table~\ref{Tab:OptValues} to their corresponding physical structures. Furthermore, as shown in Fig.~\ref{fig:UCResponse}, the unit cell does not rely on any resonances which also allows minimization of Ohmic losses.

By precisely tuning these crossed meander lines, we have sampled the optimized tensor impedance values in Table.~\ref{Tab:OptValues} and physically realized an impedance-matched CPSS as shown in Fig.~\ref{fig:CascadedStructure}. The physically-realized surface impedance values are compared with the required ones in Table.~\ref{Tab:ImpedanceComparison}.
\begin{table}[]
\begin{tabular}{cc|c|c|l|}
\cline{3-5}
 &  & Physical & \multicolumn{2}{c|}{Numerical} \\ \hline
\multicolumn{1}{|c|}{\multirow{3}{*}{Layer \#1}} & $\text{Z}_{1}^{xx}$ & $(7-j122) \Omega$ & \multicolumn{2}{c|}{$-j120 \Omega$} \\ \cline{2-5} 
\multicolumn{1}{|c|}{} & $\text{Z}_{1}^{yy}$ & $(11.13+j281) \Omega$ & \multicolumn{2}{c|}{$j280 \Omega$} \\ \cline{2-5} 
\multicolumn{1}{|c|}{} & $\text{Z}_{1}^{xy} = \text{Z}_{1}^{yx}$ & $(2.79+j247) \Omega$ & \multicolumn{2}{c|}{$j250 \Omega$} \\ \hline
\multicolumn{1}{|c|}{\multirow{3}{*}{Layer \#2}} & $\text{Z}_{2}^{xx}$ & $(4.75-j225.6) \Omega$ & \multicolumn{2}{c|}{$-j226 \Omega$} \\ \cline{2-5} 
\multicolumn{1}{|c|}{} & $\text{Z}_{2}^{yy}$ & $(6.7+j13.32) \Omega$ & \multicolumn{2}{c|}{$j10 \Omega$} \\ \cline{2-5} 
\multicolumn{1}{|c|}{} & $\text{Z}_{2}^{xy} = \text{Z}_{2}^{yx}$ & $(0.76-j90) \Omega$ & \multicolumn{2}{c|}{$-j89 \Omega$} \\ \hline
\end{tabular}
\captionsetup{justification=centering}
\caption{A comparison between the physically-realized surface impedance values from the proposed unit cells and the required values in Table \ref{Tab:OptValues}}
\label{Tab:ImpedanceComparison}
\end{table}
\noindent As seen, the reactance values of the proposed unit cells almost perfectly match to the desired ones with small resistances. The corresponding geometrical parameters of the unit cells are summarized in Table \ref{Tab:UCGeometrical}.
\begin{table}[]
\begin{tabular}{c|c|c|c|c|c|}
\cline{2-6}
                                & \textbf{$w_1$} [mm] & \textbf{$w_2$} [mm] & \textbf{$g_1$} [mm] & \textbf{$g_2$} [mm] & \textbf{$l_1$} [mm]    \\ \hline
\multicolumn{1}{|c|}{Layer \#1} & 2.313      & 1.365      & 0.08       & 0.1        & 1.658         \\ \hline
\multicolumn{1}{|c|}{Layer \#2} & 0.9362     & 1.363      & 0.1        & 0.1        & 1.271         \\ \hline
                                & \textbf{$l_2$} [mm] & \textbf{$h_1$} [mm] & \textbf{$h_2$} [mm] & \textbf{$w_a$} [mm] & \textbf{$\theta$} \\ \hline
\multicolumn{1}{|c|}{Layer \#1} & 1.329      & 0.3        & 0.3        & 0.08       & 64.4$^\circ$      \\ \hline
\multicolumn{1}{|c|}{Layer \#2} & 1.271      & 0.3        & 0.3m        & 0.08       & 18.5$^\circ$      \\ \hline
\end{tabular}
\captionsetup{justification=centering}
\caption{Physical geometries of the unit cells that implement the required impedance values in Table \ref{Tab:OptValues}}
\label{Tab:UCGeometrical}
\end{table}
\noindent Based on the cascaded physical unit cells, we have performed a full-wave simulation to verify the physically-realized impedance values. For this, the insertion loss (IL), return loss (RL), and axial ratio (AR) of the reflected/transmitted CP fields are examined. These three figure of merits are extracted by first converting the net scattering matrix obtained from the full-wave simulation (which is in an LP basis) to a scattering matrix in a CP basis as given by
\begin{equation}
\begin{aligned}
&\begin{bmatrix}
\text{S}_{11}^{\text{RR}} & \text{S}_{11}^{\text{RL}} & \text{S}_{12}^{\text{RR}} & \text{S}_{12}^{\text{RL}}\\ 
\text{S}_{11}^{\text{LR}} & \text{S}_{11}^{\text{LL}} & \text{S}_{12}^{\text{LR}} & \text{S}_{12}^{\text{LL}}\\ 
\text{S}_{21}^{\text{RR}} & \text{S}_{21}^{\text{RL}} & \text{S}_{22}^{\text{RR}} & \text{S}_{22}^{\text{RL}}\\ 
\text{S}_{21}^{\text{LR}} & \text{S}_{21}^{\text{LL}} & \text{S}_{22}^{\text{LR}} & \text{S}_{22}^{\text{LL}}
\end{bmatrix}
= \\
&\frac{1}{2}
\begin{bmatrix}
1 & -\text{j} & 0 & 0 \\ 
1 & \text{j} & 0 & 0 \\ 
0 & 0 & 1 & \text{j} \\ 
0 & 0 & 1 & -\text{j}
\end{bmatrix}
\begin{bmatrix}
\text{S}_{11}^{xx} & \text{S}_{11}^{xy} & \text{S}_{12}^{xx} & \text{S}_{12}^{xy}\\ 
\text{S}_{11}^{yx} & \text{S}_{11}^{yy} & \text{S}_{12}^{yx} & \text{S}_{12}^{yy}\\ 
\text{S}_{21}^{xx} & \text{S}_{21}^{xy} & \text{S}_{22}^{xx} & \text{S}_{22}^{xy}\\ 
\text{S}_{21}^{yx} & \text{S}_{21}^{yy} & \text{S}_{22}^{yx} & \text{S}_{22}^{yy}
\end{bmatrix}
\begin{bmatrix}
1 & 1 & 0 & 0 \\ 
-\text{j} & \text{j} & 0 & 0 \\ 
0 & 0 & 1 & 1 \\ 
0 & 0 & \text{j} & -\text{j}
\label{eq:SLPtoCP}
\end{bmatrix},
\end{aligned}
\end{equation}
\noindent where the scattering parameters in a CP basis are defined as,
\begin{equation}
\text{S}^{pq}_{ij} = \left. \frac{\text{E}_{p,i}^{\text{out}}}{\text{E}_{q,j}^{\text{in}}} \right]_{\text{E}_{\kappa\neq q,l}^{\text{in}} = 0},
\label{eq:SparamCP}
\end{equation}
\noindent such that $\text{S}^{pq}_{ij}$ is the ratio between the output CP (its handedness depends on $p$ for which it is either R or L for denoting RHCP or LHCP respectively) at port $i$ to input CP ($q$ = R or L for RHCP or LHCP respectively) at port $j$ assuming that it is the only input. Based on these scattering parameters in a CP basis, we define IL, RL, and AR of the reflected/transmitted CP fields. Specifically, the IL quantifies how well a CPSS transmits a certain CP field. Hence, it is defined as,
\begin{equation}
\text{IL} = -20\text{log}_{10}(|\text{S}^{pp}_{ij}|), i \neq j.
\label{eq:IL}
\end{equation}
\noindent On the other hand, the RL quantifies how well a CPSS reflects the other handedness. Hence, it is defined as,
\begin{equation}
\text{RL} = -20\text{log}_{10}(|\text{S}^{pp}_{ii}|).
\label{eq:RL}
\end{equation}
\noindent Lastly, the axial ratio, AR, which defines the ratio between the major and minor axes of \textcolor{black}{the} polarization ellipse for the reflected and transmitted CP fields, is given as~\cite{Liljegren2013Thesis,Lundgren2016Thesis},
\begin{equation}
\text{AR} = 20\text{log}_{10}\left( \frac{\sqrt{\sqrt{1+\kappa}+1}}{\sqrt{\sqrt{1+\kappa}-1}} \right),
\end{equation}
\noindent where $\kappa$ is given as,
\begin{equation}
\kappa = \frac{1}{4}\left( \left| \frac{\text{S}^{pq}_{ii}}{\text{S}^{qq}_{ii}} \right| - \left| \frac{\text{S}^{qq}_{ii}}{\text{S}^{pq}_{ii}} \right| \right)
\label{eq:AR}
\end{equation}

\begin{figure}[]
\centering
  \captionsetup[subfigure]{justification=centering}
  \begin{subfigure}{0.51\textwidth}
    \centering
    \includegraphics[width=\textwidth]{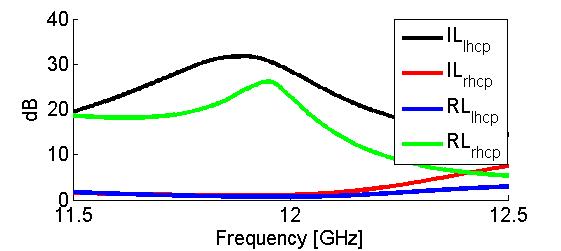}
    \caption{}
    \label{fig:ILRL}
  \end{subfigure}
  \begin{subfigure}{0.51\textwidth}
    \centering
    \includegraphics[width=\textwidth]{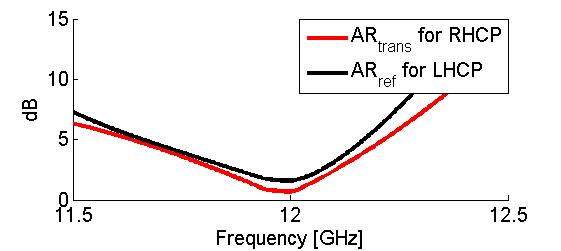}
    \caption{}
    \label{fig:AR}
  \end{subfigure}
  \caption{The full-wave simulation results. (\subref{fig:ILRL}) The insertion and return loss, and (\subref{fig:AR}) the axial ratios for the reflected LHCP field (AR$_{\text{ref}}$) and the transmitted RHCP field (AR$_{\text{trans}}$)}
  \label{fig:EvalPlot}
\end{figure}

Fig.~\ref{fig:EvalPlot} shows the full-wave simulation results for the IL, RL, and AR for the cascaded physical structure shown in Fig.~\ref{fig:CascadedStructure}.
As seen, the net scattering matrix of the cascaded unit cells closely matches to \eqref{eq:SCPSSDef} since the insertion and return losses of RHCP and LHCP fields are merely 1.15 dB and 0.69 dB, respectively, at the design frequency of 12 GHz (with all the material losses included in the simulation). On the other hand, the return loss of an RHCP field is 23.62 dB which implies that the structure is virtually reflectionless for an RHCP field (i.e., impedance-matched). In addition, the axial ratios of the reflected and transmitted CP fields are 1.56 dB and 1.25 dB, respectively. Therefore, the cascaded unit cells effectively transmits and reflects a particular handedness, while preserving the handedness of the reflected and reflected CP fields.

\subsection{Realization of spin-selective phase modulations} \label{SubsectionIIB}
Whereas the previous subsection has discussed a design of an impedance-matched CPSS, this subsection utilizes the devised impedance-matched CPSS as our building block to also demonstrate spin-selective phase modulations for re-shaping of the CP fields. To reshape the scattered CP fields, one may consider repeating the previously-demonstrated design procedure and solve for another desired scattering matrix with different $\xi$ in \eqref{eq:SCPSSDef}. However, this will impose the same phase for the reflected and transmitted CP fields. In contrast, we are interested in obtaining a spin-selective phase modulation. For this, we employ the geometric phase shift by first recalling that the cascaded unit cells can be characterized by \textcolor{black}{their} scattering matrix, $\mathbf{S}_{\text{UC}}$ in \eqref{eq:SCPSSDef} (as verified from our previous analyses), where $\text{S}_{11}^{xx}$ and $\text{S}_{11}^{yy}$ have the same magnitudes but they are exactly 180$^\circ$ apart. Such a phase relation suggests that one can induce a reflection phase shift based on the Pancharatnam-Berry phase shift. To see this point, we consider an array of the cascaded unit cells and locally rotating one of them by $\gamma$. The local output from the rotated cascaded unit cells for an incident LHCP field can be calculated as,
\begin{equation}
\begin{aligned}
\begin{bmatrix}
\text{E}_{x,1}^{\text{out}} \\ 
\text{E}_{y,1}^{\text{out}} \\ 
\text{E}_{x,2}^{\text{out}} \\ 
\text{E}_{y,2}^{\text{out}}
\end{bmatrix}
&=
\frac{1}{\sqrt{2}}
\begin{bmatrix}
\mathbf{R}(\gamma) & \mathbf{0}\\ 
\mathbf{0} & \mathbf{R}(\gamma)
\end{bmatrix}
\mathbf{S}_{\text{UC}}
\begin{bmatrix}
\mathbf{R}(\gamma) & \mathbf{0}\\ 
\mathbf{0} & \mathbf{R}(\gamma)
\end{bmatrix}^{-1}
\begin{bmatrix}
1 \\ 
j \\ 
0 \\ 
0
\end{bmatrix} \\
&=
\frac{1}{\sqrt{2}}
\begin{bmatrix}
1 \\ 
-j \\ 
0 \\ 
0
\end{bmatrix} e^{j2\gamma},
\end{aligned}
\label{eq:RotatedOutputLHCP}
\end{equation}
\noindent where $\mathbf{R}(\gamma)$ is the 2$\times$2 rotational matrix which is given in \eqref{eq:R2by2} and $\mathbf{0}$ represents a 2$\times$2 null matrix. As seen, the output is an LHCP field (note the flip of the sign for $\text{E}_{y,1}^{\text{out}}$ since the wave is traveling in the negative $z$ direction) which acquires a reflection phase shift that corresponds to exactly twice that of the physical rotation angle, $\gamma$. On the contrary, the transmitted RHCP field does not sense any physical rotation because $\text{S}_{21}^{xx}$ and $\text{S}_{21}^{yy}$ are in phase. Specifically, the output for an incident RHCP field is given as,
\begin{equation}
\begin{aligned}
\begin{bmatrix}
\text{E}_{x,1}^{\text{out}} \\ 
\text{E}_{y,1}^{\text{out}} \\ 
\text{E}_{x,2}^{\text{out}} \\ 
\text{E}_{y,2}^{\text{out}}
\end{bmatrix}
&=
\frac{1}{\sqrt{2}}
\begin{bmatrix}
\mathbf{R}(\gamma) & \mathbf{0}\\ 
\mathbf{0} & \mathbf{R}(\gamma)
\end{bmatrix}
\mathbf{S}_{\text{UC}}
\begin{bmatrix}
\mathbf{R}(\gamma) & \mathbf{0}\\ 
\mathbf{0} & \mathbf{R}(\gamma)
\end{bmatrix}^{-1}
\begin{bmatrix}
1 \\ 
-j \\ 
0 \\ 
0
\end{bmatrix} \\
&=
\frac{1}{\sqrt{2}}
\begin{bmatrix}
0 \\ 
0 \\ 
1 \\ 
-j
\end{bmatrix},
\end{aligned}
\label{eq:RotatedOutputRHCP}
\end{equation}
\noindent where it is seen that the transmitted field is still an RHCP field and it is the same as that of the case where $\gamma = 0^\circ$. Therefore, based on the theory of Pancharatnam-Berry phase shift, we can selectively apply a phase shift only for the reflected LHCP field without affecting the transmitted RHCP field and still \textcolor{black}{preserve} the impedance-matching condition. This also implies that, if one desires to also independently control the phase for the transmitted CP field, one can repeat the design procedure outlined in the previous subsection with a new value of $\xi$ in \eqref{eq:SCPSSDef} for the transmission phase shift and utilize the Pancharatnam-Berry phase for the reflection phase shift. However, for brevity, the remainder of this study considers a constant phase for the transmitted RHCP field such that the same physical cascaded unit cells can be employed where they can be simply rotated to selectively apply a certain phase shift only for the reflected CP field.

\begin{figure}[]
\centering
  \captionsetup[subfigure]{justification=centering}
  \begin{subfigure}{0.225\textwidth}
    \centering
    \includegraphics[width=\textwidth]{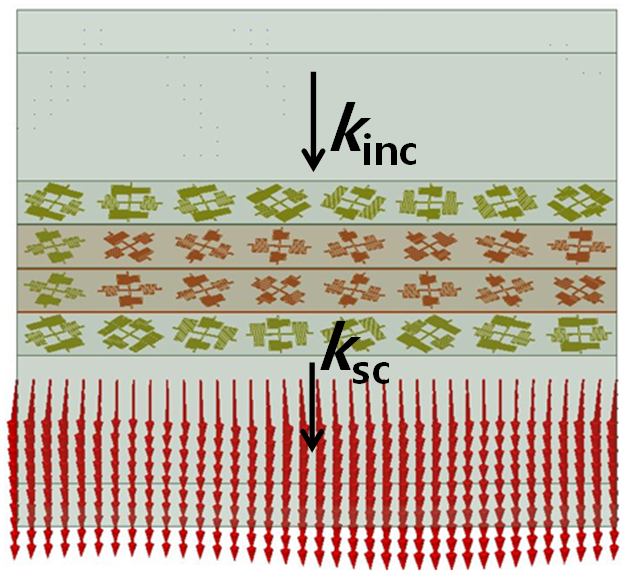}
    \caption{}
    \label{fig:RHCP}
  \end{subfigure}
  \begin{subfigure}{0.25\textwidth}
    \centering
    \includegraphics[width=\textwidth]{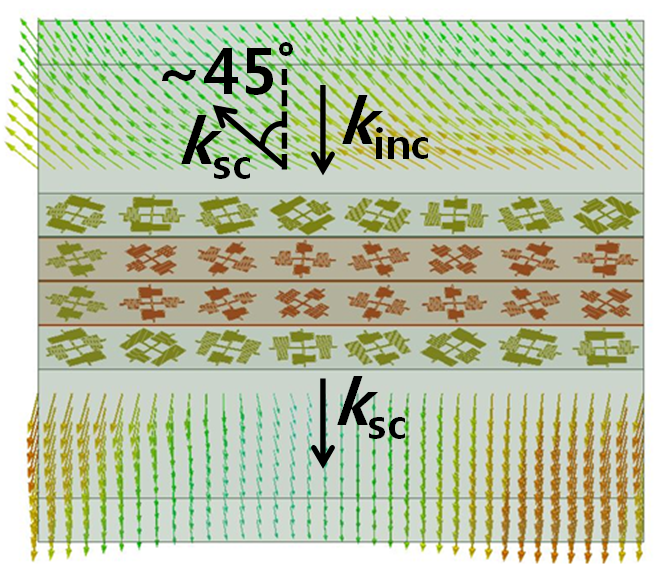}
    \caption{}
    \label{fig:LHCP}
  \end{subfigure}
  \\
  \begin{subfigure}{0.4\textwidth}
    \centering
    \includegraphics[width=\textwidth]{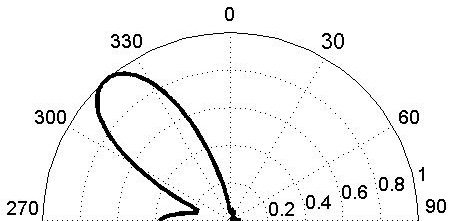}
    \caption{}
    \label{fig:FarField}
  \end{subfigure}
  \caption{The scattered Poynting vector distribution and far-field radiation plot from an array of the rotated cascaded unit cells. (\subref{fig:RHCP}) Excitation of a normally-incident RHCP field and (\subref{fig:LHCP}) LHCP field. $k_{\text{inc}}$ and $k_{\text{sc}}$ denote the wave-vectors for the incident and scattered fields, respectively. \textcolor{black}{(\subref{fig:FarField}) Far-field radiation plot of the array in case of an LHCP field excitation.}}
  \label{fig:PoyntingVector}
\end{figure}

To verify the proposed idea on the spin-selective phase modulation, we have performed a full-wave simulation based on an array of 8 cascaded unit cells that have been designed in the previous subsection (shown in Fig.~\ref{fig:CascadedStructure}). Each cascaded unit cell is progressively rotated by 22.5$^\circ$ in the lateral direction as shown in Fig.~\ref{fig:PoyntingVector} to reflect a normally-incident LHCP field at 45$^\circ$.
\noindent Fig.~\ref{fig:RHCP} shows the scattered Poynting vector distribution in case of an incident RHCP field where we see that there is virtually no reflection from the array and the field is normally transmitted. In contrast, for an incident LHCP field, \textcolor{black}{the reflected LHCP field is directed \textcolor{black}{at around} $45^\circ$ as can be verified from Figs.~\ref{fig:LHCP} and \ref{fig:FarField}.} As such, the full-wave simulation verifies that any desired reflection phases can be obtained simply by rotating the cascaded unit cells while still satisfying the impedance-matching condition for the transmitted CP field.

\section{Experimental verification}

\begin{figure}[]
\centering
  \captionsetup[subfigure]{justification=centering}
  \begin{subfigure}{0.4\textwidth}
    \centering
    \includegraphics[width=\textwidth]{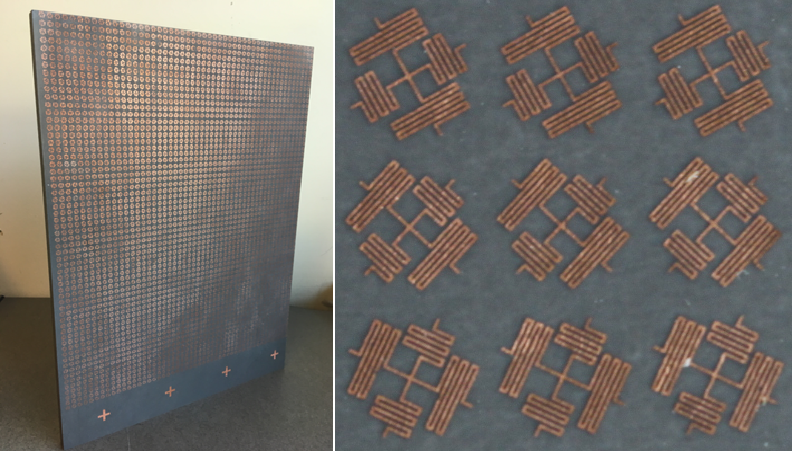}
    \caption{}
    \label{fig:FabSample}
  \end{subfigure} \\
  \begin{subfigure}{0.4\textwidth}
    \centering
    \includegraphics[width=\textwidth]{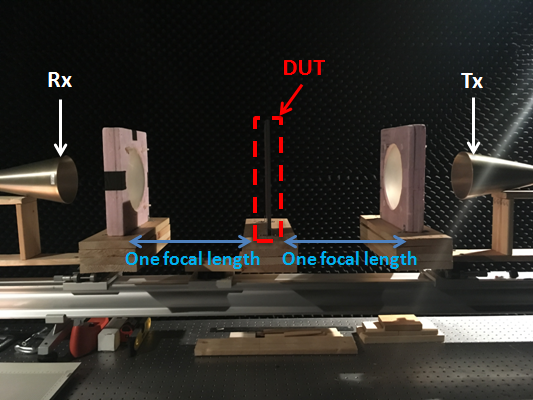}
    \caption{}
    \label{fig:QOSetup}
  \end{subfigure}
  \caption{The experimental set up for characterizing the transmission properties of the fabricated sample. (\subref{fig:FabSample}) The fabricated sample and (\subref{fig:QOSetup}) the free-space quasi-optical measurement set up}
  \label{fig:RHCPExp}
\end{figure}

To further verify the proposed idea on the spin-selective phase modulation, we have fabricated an array of the cascaded unit cells as shown in Fig.~\ref{fig:FabSample}. In particular, each cascaded unit cell in the fabricated sample is progressively rotated by 18$^\circ$ in the lateral direction from 0$^\circ$ to 360$^\circ$ to reflect a normally-incident LHCP field at 30$^\circ$ off broadside while normally transmitting an incident RHCP field. Although a rotation from 0$^\circ$ to 180$^\circ$ is sufficient for applying full 360$^\circ$ of reflection phase shift as evident from \eqref{eq:RotatedOutputLHCP}, we have deliberately rotated them from 0$^\circ$ to 360$^\circ$ to avoid sudden structural changes at the phase wraps which are known to cause the excitation of unwanted higher order Floquet modes~\cite{Lau2012DecTAP}.

\begin{figure}[]
\centering
  \captionsetup[subfigure]{justification=centering}
  \begin{subfigure}{0.45\textwidth}
    \centering
    \includegraphics[width=\textwidth]{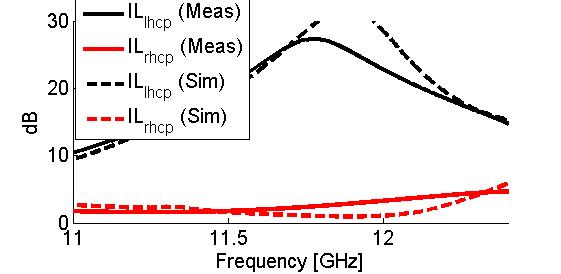}
    \caption{}
    \label{fig:ILRL_EXPT}
  \end{subfigure}
  \begin{subfigure}{0.45\textwidth}
    \centering
    \includegraphics[width=\textwidth]{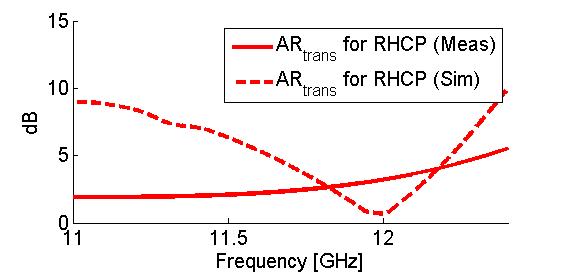}
    \caption{}
    \label{fig:AR_EXPT}
  \end{subfigure}
  \caption{The experimental results. (\subref{fig:ILRL}) The comparison between the measured (solid curves) and simulated (dotted) IL for the transmitted RHCP and LHCP fields. (\subref{fig:AR}) The measured (solid) and simulated (dotted) axial ratios of the transmitted RHCP field.}
  \label{fig:TxMeas}
\end{figure}

For measuring the transmitted RHCP field, \textcolor{black}{a} quasi-optical measurement set up has been utilized as shown in Fig.~\ref{fig:QOSetup}.
The set up consists of transmitting (Tx) and receiving (Rx) LP standard gain horn antennas with two lenses in between. The fabricated sample (DUT) sits one focal length \textcolor{black}{away} from each lens such that the phase front of the wave impinging on the sample mimics that of a plane wave. On the other hand, the output of each horn antenna is modeled as a Gaussian beam such that the optimal distance between the horn antennas is determined to be 400 mm~\cite{Goldsmith1998Wiley}. Before the measurement, the system has been calibrated based on the standard two-port TRL calibration, where a metal plate is used as the reflect standard at the location of the DUT while the reference plane of the DUT is defined as thru. To properly define a quarter-wavelength line, we have placed the antennas, lenses, and the DUT on a micrometer translation stage. Once the system has been calibrated, we have performed three separate measurements to measure the co-polarized and cross-polarized transmission coefficients (i.e., $\text{S}_{21}^{xx}$,  $\text{S}_{21}^{yx}$, and $\text{S}_{21}^{yy}$) by aligning \textcolor{black}{the} Tx and Rx antennas, and rotating one of them by 90$^\circ$ with respect \textcolor{black}{to the other}. This allows us to completely characterize the transmission properties of the fabricated sample in terms of LP waves. It should be noted that, in each measurement, we have applied time gating for all of the measured scattering parameters to filter out the unwanted reflections from the lenses and horn antennas. The measured transmission coefficients are then converted to a CP basis by utilizing \eqref{eq:SLPtoCP} and the corresponding IL, RL, and AR are computed which are compared with the simulated values in Fig.~\ref{fig:TxMeas}.
\noindent As seen, the measured IL for an RHCP field closely matches to that of the simulated values where the measured IL is 1.76 dB at 11.5 GHz. The shift in the operating frequency \textcolor{black}{may be attributed} to fabrication imperfections. Additionally, the measured IL for an LHCP field also closely matches to that of the simulated one where the measured value is 20.34 dB at 11.5 GHz. As expected, there is a large difference in the transmission of the two orthogonal CP fields. On the other hand, the AR of the transmitted RHCP field does not perfectly align with the simulated values. This \textcolor{black}{may be attributed} to any phase errors that occurred when positioning the metal plate for defining the reflect standard since any slight deviation translates to large artificial reflection and transmission phases. Nevertheless, the measured AR is still below 3 dB at 11.5 GHz (2.08 dB).

\begin{figure}[]
\centering
\resizebox{0.45\textwidth}{!}{\includegraphics{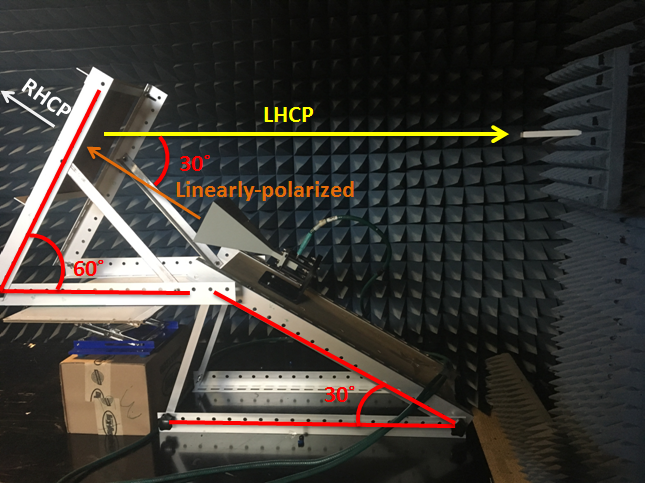}}
\caption{The near-field measurement set up}
\label{fig:NSISetup}
\end{figure}

\begin{figure}[]
\centering
  \captionsetup[subfigure]{justification=centering}
  \begin{subfigure}{0.235\textwidth}
    \centering
    \includegraphics[width=\textwidth]{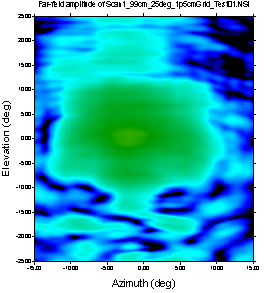}
    \caption{}
    \label{fig:RefLHCP}
  \end{subfigure}
  \begin{subfigure}{0.24\textwidth}
    \centering
    \includegraphics[width=\textwidth]{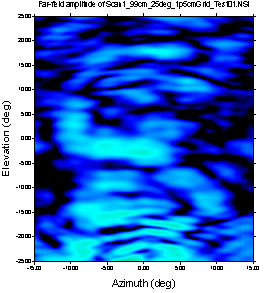}
    \caption{}
    \label{fig:RefRHCP}
  \end{subfigure}
  \begin{subfigure}{0.45\textwidth}
    \centering
    \includegraphics[width=\textwidth]{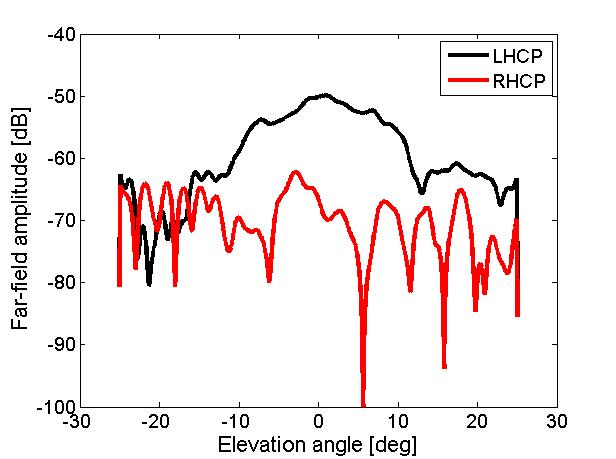}
    \caption{}
    \label{fig:RefCut}
  \end{subfigure}
  \caption{The measured far-field amplitude distributions. (\subref{fig:RefLHCP}) The reflected LHCP field and (\subref{fig:RefRHCP}) RHCP field. (\subref{fig:RefCut}) The far-field amplitude along a cut for 0$^\circ$ of azimuth angle.}
  \label{fig:RxMeas}
\end{figure}

For measuring the reflection properties of the fabricated sample, the quasi-optical measurement set up is inappropriate because an incident LHCP field is expected to be reflected at 30$^\circ$ off broadside, while all of the apparatuses are mounted on a straight micrometer translation stage. As such, we have utilized \textcolor{black}{a} near-field measurement system from NSI-MI Technologies to measure the reflected LHCP field as shown in Fig.~\ref{fig:NSISetup}.
\noindent Here, we have slanted the feed antenna at 30$^\circ$ while the fabricated sample has been tilted at 60$^\circ$ such that the feed would normally impinge to the sample and the reflected LHCP field would align with the near-field probe. The feed is \textcolor{black}{a} LP standard gain horn antenna; hence a single measurement characterizes the reflection properties of the sample for an LHCP field and \textcolor{black}{a} RHCP field. Fig.~\ref{fig:RxMeas} shows the summary of far-field amplitude distributions for the reflected LHCP and RHCP field\textcolor{black}{s} at 11.5 GHz.
\noindent As shown in Fig.~\ref{fig:RefLHCP}, near the center of the scan range, we see a distinct beam of LHCP field. On the contrary, we do not observe a clear beam for an RHCP field as shown in Fig.~\ref{fig:RefRHCP}. In particular, Fig.~\ref{fig:RefCut} shows a cut along the elevation angle at 0$^\circ$ of azimuth angle. Whereas there is a clear beam observed for the reflected LHCP field, the amplitude of the RHCP field is well below -60 dB for all elevation angles. The difference between the amplitudes of the reflected LHCP and RHCP field\textcolor{black}{s} at the center of the scan range is 20 dB. As such, the fabricated sample indeed reflects an incident LHCP field at 30$^\circ$ off broadside while transmitting most of an incident RHCP field as verified from the quasi-optical measurement.

\section{Conclusion}
This paper has proposed and experimentally demonstrated an impedance-matched CPSS which also offers spin-selective phase modulation in the microwave regime. In particular, we have cascaded four tensor impedance layers to first realize an impedance-matched CPSS. These tensor impedance layers encode particular impedance values which allow satisfying the impedance-matching condition for one handedness of a CP field, while maximizing the reflection for the opposite handedness. A numerical synthesis technique for obtaining \textcolor{black}{the required} impedance values has been discussed in detail based on the multi-conductor transmission line system. Additionally, we have proposed crossed meander lines as our unit cell for implementing the required impedance values at the operating frequency of 12 GHz. On the other hand, a spin-selective phase modulation has been realized by \textcolor{black}{realizing} a phase difference of 180$^\circ$ between the co-polarized reflection coefficients of the devised CPSS thereby leveraging the theory of Pancharatnam-Berry phase. We have experimentally demonstrated the proposed concept by utilizing a free-space quasi-optical set up and a near-field measurement system to confirm nearly-reflectionless transmission of an RHCP field and anomalous reflection of an LHCP field, respectively, \textcolor{black}{at 12 GHz}.

\section{Acknowledgment}
We would like to thank Professor Sean V. Hum for allowing us to use the near-field measurement system and Dr. Elham Baladi for her guidance in accessing the system. We also would like to thank Rogers Corporation for providing Rogers 5880 substrate samples.

\bibliographystyle{IEEEtran}
\bibliography{MinseokKim_Refs}

\end{document}